\documentclass[twocolumn,prl,aps,showpacs]{revtex4}
\usepackage{graphicx,psfig}

\newcommand{\beq}{\begin{equation}}
\newcommand{\eeq}{\end{equation}}

\begin{document}

\title{Fermi arcs and the hidden zeros of the Green's function in the pseudogap state}

\author{Tudor D. Stanescu and Gabriel Kotliar}
\affiliation{Center for Materials Theory, Department of Physics and Astronomy,
Rutgers University, Piscataway, New Jersey 08854-8019}
\date{\today}

\begin{abstract}
We investigate  the low energy properties of a correlated metal
in the proximity of a Mott insulator within the Hubbard model in
two dimensions. We introduce a new  version of the Cellular
Dynamical Mean Field Theory using  cumulants as the basic
irreducible objects. These are used for  re-constructing the
lattice quantities from their cluster counterparts. The zero
temperature one particle Green's function is characterized by the
appearance of lines of zeros, in addition to a Fermi surface
which changes topology as a function of doping. We show that
these features are intimately connected to the opening of a
pseudogap in the one particle spectrum and provide a simple
picture for the appearance of Fermi arcs.
\end{abstract}


\maketitle

The origin of the pseudogap persists as one of the leading unresolved
problems in the physics of the copper-oxide high temperature
superconductors\cite{HTCS}.
Since a lot  of the physics of  these systems arise  from short
range correlations, cluster  extensions of single site Dynamical
Mean Field Theory (DMFT)\cite{review} are ideally suited for this
problem. In fact, using such methodologies, several groups have
found\cite{psgap} that  a pseudogap, as evidenced by a
suppression of the density of states at the Fermi level, appears
near the doped Mott insulator as described  by  the Hubbard
model. This effect is caused solely by short-range correlations
and no long range order or preformed pairs need to be invoked. In
the present work we present an extension of the cluster
methodology that allows us to  identifying the emergence of lines
of zeros of the Green's function  at zero energy (i.e. the
Luttinger surface)  in addition to the quasiparticle poles (i.e.
the Fermi surface), in the proximity to the Mott transition. These
results are similar to those found in quasi one-dimensional
systems by Essler and Tsvelik\cite{tsve}. The appearance and the
evolution of a pseudogap in the particle spectral function is
governed by the topology of these lines. At small hole doping,
the Fermi surface, i.e. the line of poles, is a hole pocket having 
a Luttinger surface in close proximity.
The quasiparticle weights  along a portion of the Fermi contour
are suppressed by the proximity of the zero line, generating the  Fermi arc
behavior of the  spectral function which was identified
experimentally \cite{exper}.

We formulate a new cluster approach based on a re-summation of a
strong coupling expansion around the atomic limit, which
generalizes the CDMFT approach\cite{cdmft}. We  use the notations of
Ref.\onlinecite{paper1}, for a general lattice Hamiltonian,
\begin{eqnarray}
H &=& H_{0}+H_{1} \label{ham1} \\
&=&  \sum_{i}\sum_{\alpha }\lambda _{\alpha }X_{i}^{\alpha \alpha } + \sum_{i\neq j}\sum_{\alpha ,\beta ,\alpha ^{\prime },\beta ^{\prime
}}E_{ij}^{\alpha \beta \alpha ^{\prime }\beta ^{\prime }}X_{i}^{\alpha \beta
}X_{j}^{\alpha ^{\prime }\beta ^{\prime }},  \nonumber
\end{eqnarray}
where the local, $H_{0}$, and the non-local, $H_{1}$, terms are  expressed in
  terms of Hubbard operators, $X_{i}^{\alpha \beta }$.
Here $\alpha ,\beta ,\alpha ^{\prime }$ and $\beta ^{\prime }$
represent single-site states.  All the on-site
contributions, like, for example, the Hubbard U interaction, are included in
$\lambda_{\alpha}$, while the non-local coupling constants
$E_{ij}^{\alpha \beta \alpha ^{\prime }\beta ^{\prime }}$ can be understood
as generalized hopping matrix elements and may include hopping terms
($t_{ij}$), spin-spin interactions ($J_{ij}$), or non-local Coulomb
interactions ($V_{ij}$).

A cluster DMFT scheme\cite{cdmft} maps the lattice model onto
an effective impurity problem
 on a real space cluster ${\mathcal{C}}$, defined  by the statistical operator
\begin{eqnarray}
e^{-\beta H_{c0}}\hat{T}\exp\left\{-\int_0^{\beta} d\tau \int_0^{\beta} d\tau\prime~ X_a^{\mu}(\tau)\left[\Delta_{ab}^{\mu\nu}(\tau-\tau^{\prime}) \right. \right.\nonumber  \\
+ \left.\left.
 \delta_{\tau,\tau^{\prime}}E_{ab}^{\mu\nu}\right]X_b^{\nu}(\tau^{\prime}) +
\int_0^{\beta} d\tau~ h_a^{\mu}(\tau) X_a^{\mu}(\tau)\right\}, \label{impur}
\end{eqnarray}
where $H_{c0}= \sum_{a\in \mathcal{C}}\sum_{\alpha }\lambda
_{\alpha }X_{a}^{\alpha \alpha }$ is the local cluster
Hamiltonian, $\hat{T}$ represents the imaginary time ordering
operator, and we used the notation $(\alpha\beta)=\mu$. The
hybridization $\Delta_{ab}^{\mu\nu}$ and the effective magnetic
field $h_a^{\mu}$ are the Weiss fields describing the effects of
the rest of the system on the cluster.

A cluster DMFT approach  to  a lattice problem consists of two
elements: 1) a recipe for expressing the Weiss fields in terms of
cluster quantities , i.e. a self-consistency condition, and 2) a
recipe for determining lattice quantities from the relevant
cluster counterparts, i.e. a periodization procedure. To carry
out the first step we follow the CDMFT approach and
 construct a super-lattice by translating a cluster
$\mathcal{C}$ so as to cover the original lattice and treat the
cells as ``single'' sites with internal degrees of freedom. The
self-consistency equation for $\Delta_{ab}^{\mu\nu}$ is
determined by the condition that the Green's function for the
Hubbard operators be the same for a cell and for the impurity
cluster:
\begin{equation}
\sum_{\kappa \in RBZ} [\hat{\bf M}_c^{-1} - \hat{\bf E}_{\kappa}]^{-1} =
[\hat{\bf M}_c^{-1} - \hat{\bf \Delta} - \hat{\bf E}]^{-1},                \label{selfc}
\end{equation}
where we used the tensor notation $\hat{\bf A} = A_{ab}^{\mu\nu}$ with
$a$ and $b$ labeling the sites inside the cluster of size $N_c$,
and $\mu=(\alpha,\beta)$, $\nu=(\alpha^{\prime},\beta^{\prime})$.
 In Eq.(\ref{selfc}) $M_c$ is the irreducible
two-point cumulant of the cluster defined as the sum of all two point
diagrams generated by the strong coupling expansion of Eq. (\ref{impur})
that are irreducible with respect to $E$ and $\Delta$,
$E$ represents the coupling constant matrix  and the
$\kappa$ summation is performed over the reduced Brillouin zone
associated with a super-lattice with cells of size $N_c$. The
Weiss field $h_a^{\mu}$ is determined by
\begin{equation}
\sum_{i \notin \mathcal{C}} \sum_{\nu} E_{ai}^{\mu\nu} \langle
X_i^{\nu}\rangle = h_a^{\mu} + \sum_{b \in \mathcal{C}} \sum_{\nu}
\Delta_{ab}^{\mu\nu}(0)\langle X_b^{\nu}\rangle, \label{selfc1}
\end{equation}
where the hybridization is evaluated at zero frequency. Within
our approach, the cluster problem defined by Eq.
(\ref{impur}) has to be solved self-consistently together with
Eqs. (\ref{selfc}) and (\ref{selfc1}). Notice that only cluster
quantities, in particular the irreducible cumulant $M_c$, enter
the self-consistency loop.


The impurity model delivers cluster quantities and, to make
connection with the original lattice problem,  we need to infer
from them estimates for the lattice Green's function. A natural
way to produce these estimates is by considering the super-lattice
construction described before and averaging the relevant
quantities, which we denote below by W,  to restore periodicity,
namely
\begin{equation}
W(i-j) \approx \frac{1}{N_s}\sum_{k}~W^{SL}_{k,k+i-j},        \label{averageW}
\end{equation}
where $W$ and $W^{SL}$ are the lattice and super-lattice quantities,
respectively, and $N_s$ represents the total number of sites.
We stress that
Eq. (\ref{averageW}) represents a  super-lattice average, not a
cluster average. In particular, if W is the irreducible cumulant,
all the contributions with $k$ and $k+i-j$ belonging to different
cells are zero by construction.
One possibility\cite{AMT}, is to  periodize the Green's function
\begin{eqnarray}
\mathbf{G}({\bf k},\omega) = \frac{1}{N_c}\sum_{a,b\in\mathcal{C}}
[\hat{\bf M}_c^{-1} - \hat{\bf E}_{\bf k}]^{-1}_{ab}~e^{i{\bf
k}({\bf r}_a-{\bf r}_b)}, \label{Gk}
\end{eqnarray}
$\hat{\bf E}_{\bf k}$ being the Fourier transform of the
``hopping'' on the super-lattice, and $N_c$ the number of sites
in a cell. A second possibility, suggested by the  the strong
coupling approach investigated in this letter, is  to first
periodize the irreducible cumulant  and then use it to
reconstruct the lattice Greens function $\hat{\bf G}({\bf
k},\omega)= [\hat{\bf M}^{-1} ({\bf k},\omega ) - \hat{\bf
E}_{\bf k}]^{-1}$. For example, within a four-site approximation
(plaquette) we obtain after performing the average
(\ref{averageW}) and then taking the Fourier transform,
\begin{equation}
\mathbf{M}({\bf k},\omega) = \mathbf{M}_0(\omega) +
\mathbf{M}_1(\omega)~\alpha({\bf k}) +
\mathbf{M}_2(\omega)~\beta({\bf k}), \label{mk}
\end{equation}
where $\alpha({\bf k}) = \cos(k_x)+\cos(k_y)$,
$\beta({\bf k}) = \cos(k_x)\cos(k_y)$ and
$M_{p=\{0,1,2\}}$ represents the on-site, nearest neighbor, and
next nearest neighbor cluster cumulant, respectively.

To test the dependence of the approach on the super-lattice
construction, we also introduce an alternative self-consistency
condition that involves the periodized {\it lattice} quantities,
instead of the {\it cluster} quantities that appear
 in Eq. (\ref{selfc}),  in the spirit of PCDMFT\cite{pcdmft}
 but satisfying an  explicit cavity construction. We define the
hybridization function $\Delta$ as the sum of all the
contribution to the cluster irreducible cumulant coming from
outside the cluster and being connected to bare cumulants inside
the cluster by two ``hopping'' lines, namely:
\begin{equation}
\mathbf{\Delta}_{a b}(i\omega_n) = \sum_{A, B} ~ \mathbf{E}_{a A}
\mathbf{K}_{A B}(i\omega_n) \mathbf{E}_{B b},   \label{selfecc}
\end{equation}
where we used the matrix notation $\mathbf{W}_{ab} =
W_{ab}^{\mu\nu}$ and the matrix multiplication over $\mu$ and $\nu$
is implied.
 In Eq. (\ref{selfecc}), $\mathbf{K}_{A
B}(i\omega_n)$ represents the cavity propagator, i.e. a Green's
function which does not contain contributions arising from
irreducible cumulants having at least one site index inside a
certain cluster $\mathcal{C}$ of size $N_c$.
We assume that $\mathbf{M}_{ij}$ has a finite range $|{\bf
r}_i-{\bf r}_j|<R$, so that the terms that we subtract from the
lattice cumulants to construct the cavity form a matrix
$\mathbf{M}^*$ which is nonzero only inside an extended cluster
$\mathcal{C}_{ext}$ containing sites that can be coupled with
the  original cluster by a non-zero cumulant.
Explicitly, $\mathbf{M}_{ij}^* = \mathbf{M}_{ij}$ if at least one of the
indexes belongs to the cluster $\mathcal{C}$ and zero otherwise,
which insures that
$\mathbf{M}_{ij}^*$ is contained in $\mathcal{C}_{ext}$.
The propagator
$\mathbf{K}_{A B}(i\omega_n)$ is a generalization of the cavity
function\cite{review} and we are interested to express it in
terms of the lattice Green's function for sites $(A, B)$ that can
be connected with the cluster $\mathcal{C}$ via a hopping line.
We assume that $(A, B) \in \mathcal{C}_{ext}$, i.e. the hopping has the same
range as $\mathbf{M}_{ij}$ or smaller.
For the extended cluster the cavity propagator can be written as
\begin{equation}
\hat{\mathbf K} = \hat{\mathbf G} - \hat{\mathbf H}\hat{\mathbf
M}^{*}[\hat{\mathbf I} + \hat{\bf\mathcal{E}}\hat{\mathbf
M}^{*}]^{-1}\hat{\mathbf H},       \label{KdeG}
\end{equation}
where all the matrices ${\mathbf W}_{AB}$ are defined on the
extended cluster, $A, B \in \mathcal{C}_{ext}$, $G$ is the
lattice Green's function corresponding to the irreducible lattice
cumulant $M$, ${\mathbf H}_{AB} = ({\mathbf G}{\mathbf
M}^{-1})_{AB}$, and ${\bf\mathcal{E}}_{AB} = ({\mathbf E}{\mathbf
H}^{-1})_{AB}$.
The lattice Green's function can be expressed directly in terms of
cluster cumulants using Eq. (\ref{mk}) as
\begin{equation}
\hat{\bf G}^{-1}({\bf k},\omega)= [\mathbf{M}_0(\omega) +
\mathbf{M}_1(\omega)~\alpha({\bf k}) +
\mathbf{M}_2(\omega)~\beta({\bf k})]^{-1} - \hat{\bf E}_{\bf k}.   \label{GMk}
\end{equation}
 We note that equation (\ref{selfecc}) together with Eq.
(\ref{KdeG}) and (\ref{GMk})  can be viewed as an independent cluster scheme
with
variations that can be generated using different choices for the
cavity matrix $\mathbf{M}^*$.
This alternative  cluster method, allows us to check
that our results are self consistent by reintroducing the lattice
cumulant into the DMFT equations. This is important since previous
results of a straightforward strong coupling expansion were shown to disappear
in a more sophisticated DMFT treatment\cite{BGLG}.

\begin{figure}
\begin{center}
\includegraphics[width=0.4\textwidth]{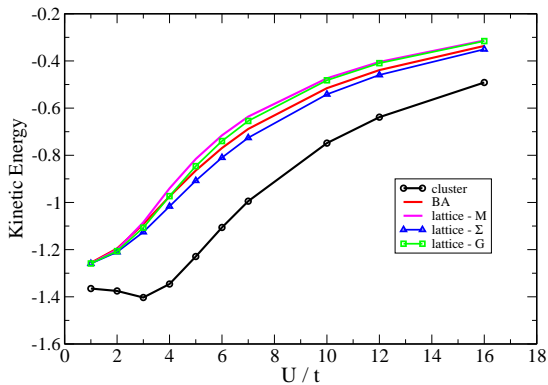}
\caption{Kinetic energy of the half-filled one-dimensional
Hubbard model as a function of the on-site interaction U at zero
temperature using: the Bethe ansatz (red line),  the  cluster
Green's function (black circles) and the lattice Green's function
obtained by periodizing  G (green squares), the self-energy (blue
triangles) and the irreducible cumulant  (magenta line).
The results are obtained using two site CDMFT and an exact
diagonalization (ED) impurity solver.}
\label{FIG1}
\end{center}
\end{figure}

We   benchmark our approach, as in ref. \cite{marcell},
by  computing the kinetic energy of the half filled one dimensional
Hubbard model
which is given known  exactly from  the Bethe ansatz.
We also make a comparison with the alternative periodization procedures
involving the Green's function and the self-energy.
Shown in Fig. (\ref{FIG1}) is the kinetic energy of the half-filled
one-dimensional Hubbard model. The exact
result from the Bethe ansatz (red) is used as a benchmark.
We notice
that the values of the kinetic energy given by the cluster Green's function
(black) are significantly different from the exact result,
while the curves obtained using the lattice Green's function, extracted using
 various procedures, cluster around the Bethe ansatz line.
 We notice that the results
obtained  by periodizing the Green's function (green) and
those obtained by periodizing the cumulant (magenta) are remarkably
similar, especially
 in the strong coupling regime.
We observed a very similar behavior in the two dimensional case.
We conclude that within CDMFT and other related cluster schemes,
when applied to small clusters,  observables should be always
extracted from the physical  {\it lattice} quantities and  {\it
not} from their {\it cluster} counterparts. Because in our
generalized strong-coupling construction of the cluster
approximations hopping is treated on equal footing with other
non-local contributions to the Hamiltonian, such as spin-spin
interaction, the conclusions derived from the calculation of  the
kinetic energy extend to all non-local physical quantities, for
example to the spin-spin correlation function. In contrast with
non-local quantities, the local physical
quantities are well approximated by their cluster values which
have to be preserved by the reconstruction schemes.
For a homogeneous cluster (as, for example, the link or the
plaquette), periodizing the Green's function automatically
satisfies this condition for all one-particle quantities as, by
construction, $G_{ii}^{latt} = G_{aa}^{c}$. The cumulant
periodization scheme also generates a local Green's function in
good agreement with $G^{c}$. However, the self-energy
scheme fails at half filling and for small doping values as it
generates spurious states in the gap.


As a first  application of our method to a strongly correlated
metal, we study  the two-dimensional Hubbard model using a
four-site cluster approximation.  In general, the lattice Green's
function can be written as
\begin{equation}
G({\bf k},\omega) = \frac{1}{\omega - r({\bf k},\omega) - i\eta({\bf k},\omega)},    \label{rk}
\end{equation}
where $\eta({\bf k},\omega)$ represents the imaginary part of the self-energy
 and $r({\bf k},\omega) = \epsilon({\bf k}) - \mu + Re\Sigma({\bf k}, \omega)$
is the energy.
In the self-energy periodization scheme
doping values,
$\Sigma({\bf k}, \omega)$
is a linear combination of the lattice self-energies given by
\begin{equation}
\mathbf{\Sigma}({\bf k},\omega) = \mathbf{\Sigma}_0(\omega) +
\mathbf{\Sigma}_1(\omega)~\alpha({\bf k}) +
\mathbf{\Sigma}_2(\omega)~\beta({\bf k}).
 \label{sigk}
\end{equation}
In the  cumulant re-construction scheme, which describes better
the system near the Mott transition, the lattice self-energy is given by
 a highly non-linear relation
\begin{eqnarray}
\Sigma({\bf k}, \omega) &=& \omega - \mu \\ \label{cumulS}
&-& \left[ \frac{\frac{1}{2}(1-\beta)}{\omega +\mu - \Sigma_A} + \frac{\frac{1}{4}(1-\alpha +\beta)}{\omega +\mu - \Sigma_B} + \frac{\frac{1}{4}(1+\alpha +\beta)}{\omega +\mu - \Sigma_C}\right]^{-1},  \nonumber
\end{eqnarray}
where $\alpha({\bf k})$ and $\beta({\bf k})$ were defined above,
 and the diagonal cluster self-energies are $\Sigma_A = \Sigma_0-\Sigma_2$ and
$\Sigma_{B(C)} = \Sigma_0\mp 2\Sigma_1 + \Sigma_2$.
Using an exact diagonalization as a CDMFT impurity solver\cite{marce} 
one finds that at zero temperature the
imaginary parts of the cluster self-energies go to zero at zero frequency.
For the real parts, on the other hand, we distinguish two regimes. At large
dopings the diagonal cluster self-energies are dominated by the local component
$\Sigma_0$ and Eq. (\ref{cumulS}) reduces in the first approximation to
Eq. (\ref{sigk}). In this regime the physics is almost local with small
corrections due to short-range correlations. All the periodization schemes
converge and the single-site DMFT represents a good first order approximation.
\begin{figure}
\begin{center}
\includegraphics[width=0.4\textwidth]{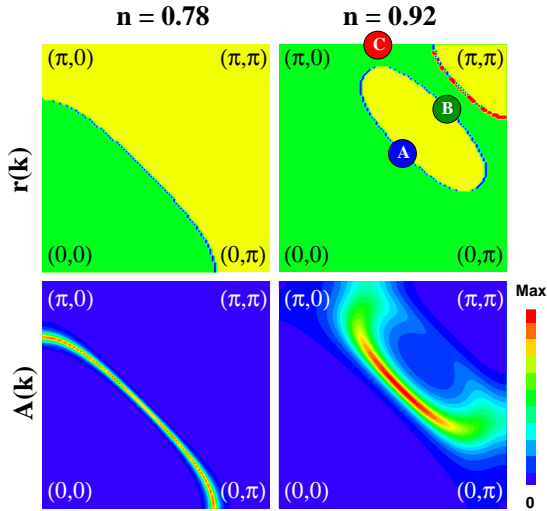}
\caption{Renormalized energy, $r({\bf k})$, (upper panels) and spectral
function, $A({\bf k})$, (lower panels) for the 2D Hubbard model with $U=8t$
and $T=0$. The color code for the upper panels is: green  ($r<0$),
blue  ($r=0$), yellow  ($r>0$), red ($r\rightarrow\infty$).}
\label{FIG2}
\end{center}
\end{figure}
In contrast, close to the Mott transition the short-range
correlations become important and the off-diagonal components of
the cluster self-energy become comparable with $\Sigma_0$. As a
consequence, at zero frequency the denominators in Eq.
(\ref{cumulS}) may acquire opposite signs generating a divergence
in the lattice self-energy. This  pole of $\Sigma({\bf k},
\omega=0)$, or equivalently of $r({\bf k})$, gives rise  to a
zero of the lattice Green's function. We show in Fig. \ref{FIG2}
the renormalized energy, $r({\bf k})$, and the spectral function
$A({\bf k}, \omega=0) = -1/\pi Im G({\bf k},0)$ for a
two-dimensional Hubbard model with $U=8t$ at zero temperature for
two values of doping. For $n=0.78$ (left panels) we have  a large
electron-type Fermi surface (blue line in the $r({\bf k})$ panel)
separating the occupied region of the Brillouin zone (green),
defined by $r({\bf k}) < 0$ from the unoccupied region (yellow)
defined by  $r({\bf k}) > 0$. The Fermi surface can be also
traced in the $A({\bf k})$ panel as the maximum of the spectral
function. On the other hand, for $n=0.92$ a qualitatively
different picture emerges. The Fermi surface (blue line) is now
represented by a hole pocket and, in addition, we have a line of
zeros of the Green's function (red line) close to the $(\pi,\pi)$
region of the Brillouin zone. Furthermore, there is no one-to-one
correspondence between the Fermi surface and the maximum of the
spectral function. This behavior has two origins: 1) the
proximity of a zero line suppresses the weight of the
quasiparticle on the far side of the pocket, and 2) for k-points
corresponding to $r({\bf k}) \neq 0$ the quasiparticles are
pushed away from $\omega=0$ and a pseudogap opens at the Fermi
level. We show this explicitly in Fig. \ref{FIG3} by comparing
the low frequency dependence of the spectral function in three
different points of the Brillouin zone, marked by A, B and C in
Fig. \ref{FIG2}. Notice the suppression of the zero frequency peak
at point B and the frequency shift $\delta = - 0.05t$ of the peak
at point
 C. The cumulant approach provides a simple interpretation of this effect,
observed in photo emission experiments\cite{exper}, in terms of
the emergence of infinite self-energy lines or  equivalently
Luttinger lines ( lines of zeros of the Greens function).

In conclusion, our strong coupling CDMFT study of the Hubbard model shows that
the lightly doped system is characterized by a small, {\it closed} Fermi
line that {\it appears} in the zero frequency spectral function  as
an {\it arc} due to the presence
of a line of zeros of the Green's function near the ``dark side'' of the Fermi
surface. These lines appear near the Mott insulator in order to satisfy
the generalized Luttinger theorem\cite{dzy},
\begin{equation}
\frac{N}{V} = 2\int_{{\mbox{Re}}G({\bf k}, \omega=0) > 0}\frac{d^2{\bf k}}{(2\pi)^2}.
\end{equation}
While this theorem is not exactly  satisfied in our finite
cluster calculation  due to the  small cluster approximation,
even this small cluster size has all the qualitative elements
needed to interpret the physics of the underdoped regime which is
controlled  by the position of the lines of zeros and infinities.
Notice that  the standard Luttinger theorem will be {\bf always}
violated in a system characterized by a pseudogap, and it is
strongly violated in our solution as the area contained in the
pockets shrinks. The  tendency to increase   the area contained by
the Luttinger surface as the pockets shrink is correctly captured
by our approximation. Furthermore, we identify the pseudogap
which is seen in leading edge study of photo emission
experiments  as the small negative shift of the spectral weight
in points of the Brillouin zone that are not on the Fermi line
(for example point C in Fig. \ref{FIG2}), which is distinct from
the larger gap between the peaks above and below the Fermi level
(see Fig. \ref{FIG3}).
\begin{figure}
\begin{center}
\includegraphics[width=0.4\textwidth]{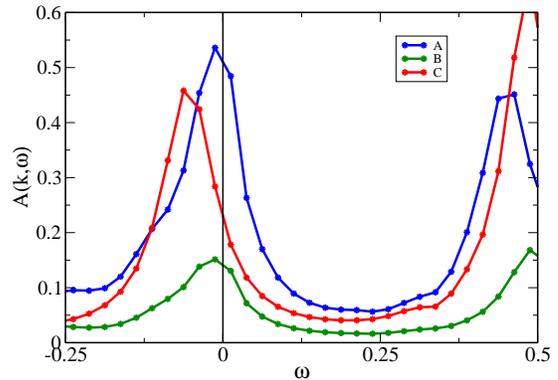}
\caption{Frequency dependence of the spectral function for three points in
the Brillouin zone marked by A, B, and C in Fig. \ref{FIG2}.}
\label{FIG3}
\end{center}
\end{figure}
Remarkably,  the  lines of poles of the self energy appear first
far from the Fermi surface. This is   a strong coupling instability
which has no weak coupling precursors on the Fermi surface. Our
results raise an interesting question. If the evolution
in Figure \ref{FIG2} from large to small doping  is continuous, it has to
go through a  critical point where the topology of the Fermi surface
(and perhaps that of the lines where the self energy is infinite) changes.
This topological change and its possible connection to an
underlying critical point at finite doping in the cuprate phase
diagram deserves  further  investigation.

ACKNOWLEDGMENTS: This research was supported by the NSF under grant DMR-0096462
and by the Rutgers University Center for Materials Theory. Many useful
discussions with M. Civelli,  B. Kyung,  A. M. Tremblay, M. Capone,
O. Parcollet and  A. Tsvelik are
gratefully acknowledged. Special thanks to K. Haule and M. Civelli
for allowing the use of their ED and NCA programs.


\begin{thebibliography}{99}

\bibitem{HTCS} T. Timusk and B Statt, Rep. Prog. Phys. {\bf 62}, 61 (1999).

\bibitem{review} A. Georges, G. Kotliar, W. Krauth, and M. J. Rozenberg, Rev.
Mod. Phys. \textbf{68}, 13 (1996).

\bibitem{psgap} T. Maier, M. Jarrel, T. Pruschke and J. Keller, Eur. Phys. J. {\bf B 13}, 613 (2000); M. Jarrel, T. Maier, M. H. Hettler, and A. N. Tahvildar-Zadeh, Europhys. Lett. {\bf 56}, 563 (2001); T.D. Stanescu and P. Philips, Phys. Rev. Lett. {\bf 91}, 017002 (2003); D. S\'en\'echal and  A.-M.S. Tremblay, Phys. Rev. Lett. 92, 126401 (2004); B. Kyung, S.S. Kancharla, D. S\'en\'echal, A.-M.S. Tremblay, M. Civelli, and G. Kotliar, cond-mat/0502565 (2005).

\bibitem{paper1} T.D. Stanescu and G. Kotliar, Phys. Rev. {\bf B 70}, 205112 (2004).

\bibitem{cdmft} G. Kotliar, S.Y. Savrasov, G. Palsson, and G. Biroli,
Phys. Rev. Lett. {bf 87}, 186401 (2001).

\bibitem{tsve} F.H.L. Essler and A.M. Tsvelik, Phys. Rev. {\bf B 65}, 1151171 (2002).

\bibitem{AMT} D. S\'en\'echal, D. Perez, and M. Pioro-Ladri\`ere, Phys. Rev. Lett. {\bf 84}, 522 (2000).

\bibitem{pcdmft} G. Biroli, O. Parcollet, and G. Kotliar, Phys. Rev. {\bf B 69}, 205108 (2004).

\bibitem{BGLG} S. Biermann, A. Georges, A. Lichtenstein, and T. Giamarchi, Phys.Rev.Lett {\bf 87}, 276405 (2001).

\bibitem{marcell} M. Capone, M. Civelli, S.S. Kancharla, C. Castellani, and G. Kotliar, Phys. Rev. {\bf B 69}, 195105 (2004).

\bibitem{marce} M. Civelli, M. Capone, S.S. Kancharla, O. Parcollet, and G. Kotliar, cond-mat/0411696.

\bibitem{exper} D.S. Marshall et al., Phys. Rev Lett. {\bf 76}, 4841 (1996); M.R. Norman et al., Nature {\bf 392}, 157 (1998); A. Damascelli, Z.X. Shen, and Z. Hussain, Rev. Mod. Phys. {\bf 75}, 473 (2003).

\bibitem{dzy} I. Dzyaloshinskii, Phys. Rev. {\bf B 68}, 085113 (2003).

\end{thebibliography}
\end{document}